\documentclass[12pt,preprint]{aastex}
 
\newcommand{\etal}{et~al.\ }
\newcommand{\PVdblt}{{\rm P}\kern 0.1em{\sc v}~$\lambda\lambda 1117, 1128$}
\newcommand{\CaIIdblt}{{\rm Ca}\kern 0.1em{\sc ii}~$\lambda\lambda 3934, 3969$}
\newcommand{\AlIIIdblt}{{\rm Al}\kern 0.1em{\sc iii}~$\lambda\lambda 1854, 1862$}
\newcommand{\CIVdblt}{{\rm C}\kern 0.1em{\sc iv}~$\lambda\lambda 1548, 1550$}
\newcommand{\MgIIdblt}{{\rm Mg}\kern 0.1em{\sc ii}~$\lambda\lambda 2796, 2803$}
\newcommand{\NVdblt}{{\rm N}\kern 0.1em{\sc v}~$\lambda\lambda 1238, 1242$}  
\newcommand{\SVIdblt}{{\rm S}\kern 0.1em{\sc vi}~$\lambda\lambda 933, 944$} 
\newcommand{\OVIdblt}{{\rm O}\kern 0.1em{\sc vi}~$\lambda\lambda 1031, 1037$} 
\newcommand{\SiIIdblt}{{\rm Si}\kern 0.1em{\sc ii}~$\lambda\lambda 1190, 1193$} 
\newcommand{\SiIVdblt}{{\rm Si}\kern 0.1em{\sc iv}~$\lambda\lambda 1393, 1402$} 
\newcommand{\PV}{\hbox{{\rm P}\kern 0.1em{\sc v}}}
\newcommand{\AlI}{\hbox{{\rm Al}\kern 0.1em{\sc i}}}
\newcommand{\AlII}{\hbox{{\rm Al}\kern 0.1em{\sc ii}}}
\newcommand{\AlIII}{{\hbox{\rm Al}\kern 0.1em{\sc iii}}}
\newcommand{\CaII}{\hbox{{\rm Ca}\kern 0.1em{\sc ii}}}
\newcommand{\CII}{\hbox{{\rm C}\kern 0.1em{\sc ii}}}
\newcommand{\CIIe}{\hbox{{\rm C$^{\ast}$}\kern 0.1em{\sc ii}}}
\newcommand{\CIII}{\hbox{{\rm C}\kern 0.1em{\sc iii}}}
\newcommand{\CIV}{\hbox{{\rm C}\kern 0.1em{\sc iv}}}
\newcommand{\CV}{\hbox{{\rm C}\kern 0.1em{\sc v}}}
\newcommand{\HI}{\hbox{{\rm H}\kern 0.1em{\sc i}}}
\newcommand{\HII}{\hbox{{\rm H}\kern 0.1em{\sc ii}}}
\newcommand{\Lya}{\hbox{{\rm Ly}\kern 0.1em$\alpha$}}
\newcommand{\Lyb}{\hbox{{\rm Ly}\kern 0.1em$\beta$}}
\newcommand{\Lyg}{\hbox{{\rm Ly}\kern 0.1em$\gamma$}}
\newcommand{\Lyd}{\hbox{{\rm Ly}\kern 0.1em$\delta$}}
\newcommand{\Lye}{\hbox{{\rm Ly}\kern 0.1em$\epsilon$}}
\newcommand{\Lyphi}{\hbox{{\rm Ly}\kern 0.1em$\phi$}}
\newcommand{\Lyfive}{\hbox{{\rm Ly}\kern 0.1em$5$}}
\newcommand{\Lysix}{\hbox{{\rm Ly}\kern 0.1em$6$}}
\newcommand{\Lyseven}{\hbox{{\rm Ly}\kern 0.1em$7$}}
\newcommand{\Lyeight}{\hbox{{\rm Ly}\kern 0.1em$8$}}
\newcommand{\Lynine}{\hbox{{\rm Ly}\kern 0.1em$9$}}
\newcommand{\Lyten}{\hbox{{\rm Ly}\kern 0.1em$10$}}
\newcommand{\Lyeleven}{\hbox{{\rm Ly}\kern 0.1em$11$}}
\newcommand{\HeI}{\hbox{{\rm He}\kern 0.1em{\sc i}}}
\newcommand{\HeII}{\hbox{{\rm He}\kern 0.1em{\sc ii}}}
\newcommand{\FeI}{\hbox{{\rm Fe}\kern 0.1em{\sc i}}}
\newcommand{\FeII}{\hbox{{\rm Fe}\kern 0.1em{\sc ii}}}
\newcommand{\FeIII}{\hbox{{\rm Fe}\kern 0.1em{\sc iii}}}
\newcommand{\MnII}{\hbox{{\rm Mn}\kern 0.1em{\sc ii}}}
\newcommand{\MgI}{\hbox{{\rm Mg}\kern 0.1em{\sc i}}}
\newcommand{\MgII}{\hbox{{\rm Mg}\kern 0.1em{\sc ii}}}
\newcommand{\MgIII}{\hbox{{\rm Mg}\kern 0.1em{\sc iii}}}
\newcommand{\NI}{\hbox{{\rm N}\kern 0.1em{\sc i}}}
\newcommand{\NII}{\hbox{{\rm N}\kern 0.1em{\sc ii}}}
\newcommand{\NIII}{\hbox{{\rm N}\kern 0.1em{\sc iii}}}
\newcommand{\NV}{\hbox{{\rm N}\kern 0.1em{\sc v}}}
\newcommand{\OVI}{\hbox{{\rm O}\kern 0.1em{\sc vi}}}
\newcommand{\OI}{\hbox{{\rm O}\kern 0.1em{\sc i}}}
\newcommand{\OII}{\hbox{[{\rm O}\kern 0.1em{\sc ii}]}}
\newcommand{\OIV}{\hbox{{\rm O}\kern 0.1em{\sc iv}]}}
\newcommand{\SI}{{\rm S}\kern 0.1em{\sc i}}
\newcommand{\SIV}{{\rm S}\kern 0.1em{\sc iv}}
\newcommand{\SVI}{{\rm S}\kern 0.1em{\sc vi}}
\newcommand{\SiI}{\hbox{{\rm Si}\kern 0.1em{\sc i}}}
\newcommand{\SiII}{\hbox{{\rm Si}\kern 0.1em{\sc ii}}}
\newcommand{\SiIII}{\hbox{{\rm Si}\kern 0.1em{\sc iii}}}
\newcommand{\SiIV}{\hbox{{\rm Si}\kern 0.1em{\sc iv}}}
\newcommand{\SII}{\hbox{{\rm S}\kern 0.1em{\sc ii}}}
\newcommand{\SIII}{\hbox{{\rm S}\kern 0.1em{\sc iii}}}
\newcommand{\NaI}{\hbox{{\rm Na}\kern 0.1em{\sc i}}}
\newcommand{\TiII}{\hbox{{\rm Ti}\kern 0.1em{\sc ii}}}
\newcommand{\kms}{\hbox{km~s$^{-1}$}}
\newcommand{\cmsq}{\hbox{cm$^{-2}$}}
\newcommand{\cc}{\hbox{cm$^{-3}$}}
 
\tighten
\begin{document}
 
 
\shortauthors{CHARLTON ET~AL.}
\shorttitle{WEAK MGII ABSORBERS TOWARD PG1634+706}


\title{High Resolution STIS/HST and HIRES/Keck Spectra of Three Weak {\MgII}
Absorbers Toward PG~1634+706\altaffilmark{1,2}}

\author{Jane~C.~Charlton\altaffilmark{3}}
\affil{Department of Astronomy and Astrophysics, The Pennsylvania State University, University Park, PA 16802; charlton@astro.psu.edu}

\author{Jie~Ding}
\affil{Department of Astronomy and Astrophysics, The Pennsylvania State University, University Park, PA 16802; ding@astro.psu.edu}

\author{Stephanie~G.~Zonak}
\affil{Department of Astronomy, University of Maryland, College Park, MD 20742-2421; szonak@astro.umd.edu}

\author{Christopher~W.~Churchill\altaffilmark{4}}
\affil{Department of Astronomy and Astrophysics, The Pennsylvania State University, University Park, PA 16802; ding@astro.psu.edu}

\author{Nicholas A. Bond}
\affil{Princeton University Observatory, Peyton Hall, Princeton, NJ 08544-0001; nbond@astro.princeton.edu}

\and

\author{Jane~R.~Rigby}
\affil{Steward Observatory, University of Arizona, Tucson, AZ, 85721; jrigby@as.arizona.edu}

\altaffiltext{1}{Based in part on observations obtained at the
W.~M. Keck Observatory, which is operated as a scientific partnership
among Caltech, the University of California, and NASA. The Observatory
was made possible by the generous financial support of the W. M. Keck
Foundation.}
\altaffiltext{2}{Based in part on observations obtained with the
NASA/ESA {\it Hubble Space Telescope}, which is operated by the STScI
for the Association of Universities for Research in Astronomy, Inc.,
under NASA contract NAS5--26555.}
\altaffiltext{3}{Center for Gravitational Physics and Geometry}
\altaffiltext{4}{Visiting Astronomer at the W.~M. Keck Observatory}

\begin{abstract}
High resolution optical (HIRES/Keck) and UV (STIS/HST) spectra,
covering a large range of chemical transitions, are analyzed for three
single--cloud weak {\MgII} absorption systems along the line of sight
toward the quasar PG~$1634+706$.  Weak {\MgII} absorption lines in
quasar spectra trace metal--enriched environments that are rarely
closely associated with the most luminous galaxies ($>0.05L^*$).  The
two weak {\MgII} systems at $z=0.81$ and $z=0.90$ are constrained to
have $\ge$~solar metallicity, while the metallicity of the $z=0.65$ system
is not as well--constrained, but is consistent with
$>1/10$th solar.  These weak {\MgII} clouds are
likely to be local pockets of high metallicity in a lower metallicity
environment.
All three systems have two phases of gas, a higher density region
that produces narrower absorption lines for low ionization transitions,
such as {\MgII}, and a lower density region that produces broader
absorption lines for high ionization transitions, such as {\CIV}.
The {\CIV} profile for one system (at $z=0.81$) can be fit with a single broad
component ($b\sim10$~{\kms}), but those for the other two systems
require one or two additional offset high ionization clouds.
Two possible physical pictures for the phase structure are discussed: one
with a low--ionization, denser phase embedded in a lower density surrounding
medium, and the other with the denser clumps surrounding more highly
ionized gas.
\end{abstract}

\keywords{quasars--- absorption lines; galaxies--- evolution;
galaxies--- halos}

\section{Introduction}
\label{sec:intro}

Weak {\MgII} absorbers, those with $W_r(2796) < 0.3$~{\AA}, constitute
$65$\% of the total {\MgII} absorber population \citep{weak1} at $z \sim 1$.
They account for a fair fraction of the $N({\HI}) \sim
10^{16}$~{\cmsq} {\Lya} forest \citep{weak2}.  Unlike strong {\MgII}
absorbers (which almost always are associated with a $>0.05L^*$ galaxy
\citep{bb91,sdp94,s95}),
weak {\MgII} absorbers can usually not be associated with a $>0.05L^*$
galaxy within impact parameter $\simeq 50 h^{-1}$~kpc of the QSO
(\citet{weak2}; C. Steidel, private communication).
This lack of an association with bright galaxies suggests that weak {\MgII}
absorbers may be a physically different population than strong {\MgII} absorbers.

Studying the physical conditions of weak {\MgII} absorbers (eg.,
metallicity, ionization conditions, total column density, and size) is
useful for two reasons: 1) Physical conditions provide clues as to the
nature of these absorbers, whose physical origin is not known; 2) Weak
{\MgII} clouds provide an opportunity to study metal--enriched
environments over a range of redshifts.  They trace metal production
either in intergalactic space or in dwarf or low surface brightness
galaxies.

A study of weak {\MgII} absorbers at $z\sim1$ requires spectra in both
the optical and in the UV in order that the several key low and high
ionization transitions (especially {\MgII}, {\FeII}, {\CIV}), as well
as the Lyman series, are covered.  Using the Keck/High Resolution
Spectrograph (HIRES) \citep{vogt94} at high resolution ($R=45,000$)
and the Faint Object Spectrograph (FOS)/{\it Hubble Space Telescope}
(HST) at low resolution ($R=1300$), \citet{weak2}
applied photoionization models to $15$ single--cloud weak {\MgII}
absorbers.  They argued that many multiple cloud weak absorbers (which
comprise $35$\% of the weak absorbers) are likely to be part of the
same population as the strong absorbers, but that single--cloud weak
systems are likely to be a different population.

At least half of the weak, single--cloud systems of \citet{weak2}
require a second phase of gas in order to reproduce the {\CIV}
absorption and/or to fit the {\Lya} profile without exceeding the
{\HI} column density derived from the Lyman limit break.  While the
{\MgII} clouds typically have Doppler parameters of $2$--$6$~{\kms},
the second phase must have a larger effective Doppler parameter ($\sim
10$--$30$~{\kms}).  With only the low resolution UV data available
to \citet{weak2}, it was not clear if this implies a single broad
component or multiple, blended clouds.  Also, in the systems for which
the second phase is not required (such as when only a limit for {\CIV}
can be derived), it was not clear if the phase is just less extreme in
its properties or if it is absent.

By comparing the {\MgII} column densities to the {\Lya} profiles,
\citet{weak2} inferred that weak, single cloud
{\MgII} absorbers have metallicities of at least one--tenth solar in
the phase of gas in which the {\MgII} absorption arises.  The sizes
and densities of most of the weak, single--cloud systems were
unconstrained; however, three of them had $N({\MgII}) \sim
N({\FeII})$, which implies low ionization conditions, relatively high
densities, $n_H = 0.1$~{\cc}, and small sizes, $\sim 10$~pc.  Those
systems with limits on {\FeII} may be part of a ``continuum'' of
single--cloud weak {\MgII} absorbers, with some having {\FeII} just
below the detection threshold, implying a continuous distribution of
ionization conditions.  Alternatively, some of the systems without
detected {\FeII} could be part of a different population, with
ionization conditions, metallicities, and/or environments distinctly
different from systems with detected {\FeII}.

In May and June 2000 high resolution UV spectra ($R=30,000$) of the
$z=1.335$ quasar, PG~$1634+706$, became public in the HST archive.  We
have high resolution optical (HIRES/Keck) spectra of this same quasar.
For the first time, weak {\MgII} absorbers can be studied through
simultaneous high resolution coverage of numerous transitions.  This
allows more direct inferences of the physical properties of the gas,
such as metallicity and phase structure.  In particular, it should
enable a determination of the nature of the second, broad phase
required for some absorbers by the large relative strength of {\CIV}
absorption in lower resolution data.  Is this higher--ionization,
broad phase centered on the {\MgII} cloud or is it offset? Is it
produced by multiple clouds or by a single, smooth structure?  Are
systems for which {\CIV} was not detected at low resolution
fundamentally different, or do they merely have a weaker second phase?
The kinematics and physical properties of the second phase, and its
relationship to the {\MgII} cloud phase, are important diagnostics of
the type of structure responsible for the weak {\MgII} absorption.

In anticipation of the release of the high resolution STIS/HST
spectra, we previously pursued an in--depth study of four {\MgII}
absorbers along the PG~$1634+706$ line of sight \citep{low1634} using
the available low resolution FOS/HST spectra and the HIRES/Keck
spectra.  Since only the brightest quasars can, in the near future,
feasibly be studied at high resolution in the UV, we aimed to compare
inferences drawn on the basis of low resolution spectra to those that
would be obtained once the higher resolution spectra were released.
Confirmation of our conclusions would lend credibility to larger
statistical studies that rely on a combination of high and low
resolution spectra \citep{weak2,archive2}.

The present paper focuses on three single--cloud weak {\MgII}
absorbers along the PG~$1634+706$ line of sight.  Two of these
absorbers, at $z=0.8182$ and $z=0.9056$, were detected in the
HIRES/Keck spectra at the sensitivity of our original weak {\MgII}
survey ($5\sigma$ for {\MgII}~$2796$) \citep{weak1}.  These two
systems are dramatically different from each other in that the
$z=0.8182$ system has a stringent limit on {\CIV} ($W_r(1548) <
0.07$~{\AA} at $3\sigma$) while the $z=0.9056$ absorber has {\CIV}
detected at $W_r(1548) = 0.18$~{\AA}.  In the latter system, the
{\CIV} arises not in the {\MgII} cloud phase, but in a second phase of
gas.  This provides an excellent opportunity both to study a system
with very weak {\CIV} and to address the nature of the second phase.
The metallicity could not be derived for the $z=0.8182$ system because
the FOS/HST spectrum did not cover any of the Lyman transitions;
however, the new STIS/HST spectra can address metallicity for this
system.  Models of the $z=0.9056$ absorber did, however, imply a
super--solar metallicity and/or a depleted or $\alpha$--enhanced
abundance pattern.  This conclusion will be re--assessed in this
paper.  Also, as we will describe in \S~\ref{sec:data}, we have now
identified a third single--cloud weak {\MgII} system at $z=0.6534$
that was just below the detection threshold of our previous study.
That system was not modeled in \citet{low1634},
but will be considered in the present paper.

We begin in \S~\ref{sec:data} by presenting, for the three single
cloud weak {\MgII} systems toward PG~$1634+706$, high resolution
profiles of all relevant transitions covered by STIS/HST or HIRES/Keck
spectra.  In \S~\ref{sec:techniques}, we discuss our strategy to infer
the physical conditions of these systems by applying Cloudy
photoionization models.
We also consider the possibility that collisional ionization dominates.
The results inferred for the phase structure,
ionization parameters/densities, metallicities, kinematics, and
abundance patterns of the three systems are presented in
\S~\ref{sec:results}.  In \S~\ref{sec:caveats}, the effect of relaxing
assumptions of abundance pattern and the shape of
the ionizing spectrum are considered.
The three weak {\MgII} systems are compared in
\S~\ref{sec:discussion}.  In that discussion, we particularly focus on
the nature of the second phase and compare the results to those
obtained on the basis of just the lower resolution FOS/HST spectrum.
Finally, we summarize our conclusions in
\S~\ref{sec:conclude}.

\section{Observations of the Three Weak {\MgII} Absorbers Toward PG~$1634+706$}
\label{sec:data}

We briefly describe the observations with HIRES/Keck, STIS/HST, and
FOS/HST of PG~$1634+706$, and then present the three single--cloud
weak {\MgII} absorbers along this line of sight at $z=0.8181$,
$z=0.9056$, and $z=0.6534$.  A WFPC2/HST image of the quasar field
exists (program 6740, S. Oliver, P. I.)  but the quasar is quite
bright so it is not possible to perform adequate PSF subtraction in
order to detect galaxies close to the line of sight.  Also, without
redshifts of candidate galaxies in the field, we could not separate
out the identities of the three weak and two strong {\MgII} absorbers.

\subsection{HIRES/Keck}
The HIRES/Keck observations covered $3723$~{\AA} to $6186$~{\AA} at
$R=45,000$ (FWHM $\sim 6.6$~{\kms}) ($3$ pixels per resolution
element).  {\MgII}, {\MgI}, and {\FeII} transitions are covered for
the three systems, but in all cases only {\MgII} is detected (at
$>3\sigma$).  Equivalent widths and limits are listed in Table~\ref{tab:ewtab}.
The spectra were obtained on 1995 July 4 and 5, and the
combined spectrum has $S/N \sim50$ over most of the wavelength range,
but gradually falling toward the lowest wavelengths.  The most
stringent limits on {\FeII} come from the reddest of the {\FeII}
transitions, at $2600$~{\AA}.  The reduction of the spectrum,
continuum fits, line identification, and procedure for Voigt profile
fitting were described in \citet{strong1}.

\subsection{STIS/HST}
The STIS/HST observations provided useful coverage from $1880$~{\AA}
to $3118$~{\AA} at $R=30,000$ (FWHM $\sim 10$~{\kms}) ($2$ pixels per
resolution element).  Two sets of echelle spectra were obtained with
two different tilts of the E230M grating, using the $0.2$~{\arcsec}
$\times$ $0.2$~{\arcsec} slit.  The first, with central wavelength
$2269$~{\AA}, was obtained by Burles \etal in 1999 May and June
(proposal ID 7292).  The total exposure time was $29,000$~s.  The
second, with central wavelength $2707$~{\AA}, was obtained by Jannuzi
\etal in 1999 June (proposal ID 8312), with a total exposure time
$26,425$~s.  The two spectra overlap in the region
$2275$--$2708$~{\AA}.  We co--added the spectra, weighted by
the exposure times, and also combined
multiple order coverage in the same spectrum.
However, we also considered the differences between
the two realizations as an indication of 
systematic errors (due to continuum fitting, correlated
noise, and unknown factors) when comparing model profiles to the data.
Reductions were done with the standard STIS pipeline and continuum
fits using the {\sc sfit} task in IRAF\footnote{IRAF is distributed by
the National Optical Astronomy Observatories, which are operated by
AURA, Inc., under contract to the NSF.}.

\subsection{The Three Weak Single Cloud {\MgII} Systems}

Based upon our original analysis of the HIRES/Keck spectrum there were
two single cloud, weak {\MgII} systems along the PG~$1634+706$ line of
sight, at $z=0.8181$ and $z=0.9056$.  After analyzing the
STIS/HST spectra, another slightly weaker {\MgII} system was found in
the HIRES/Keck spectra, just below the threshold of our previous
survey \citep{weak1}.
First, the STIS spectra were searched for {\CIV} doublets.
Only one {\CIV} doublet was found with comparable equivalent width to
those associated with the two known {\MgII} absorption systems.  The
new system, at $z=0.6534$, also had detected {\Lya}, a {\SiIV} doublet,
{\CII}, and {\SiII} in the STIS spectrum.  Based upon the equivalent
width of {\SiII} we expected that {\MgII} should be detected in the
HIRES/Keck spectrum.  We then searched that location in the HIRES
spectrum and found the {\MgII} doublet.  The $\lambda2796$ transition
was detected at the $6.8\sigma$ level, and the $\lambda2803$
transition at the $2.9\sigma$ level.  (Our previous survey had a
$3\sigma$ detection threshold for the $\lambda2803$ transition.)

Figure~\ref{fig:data81} presents detected transitions and limits of
interest in constraining the conditions in the $z=0.8181$ system.
Figure~\ref{fig:data90} displays the same for the $z=0.9056$ system,
and Figure~\ref{fig:data65} for the $z=0.6534$ system.
Table~\ref{tab:ewtab} lists the equivalent widths for detected
transitions and selected equivalent width limits ($3\sigma$) for the
three systems.  We fit each transition with the minimum
number of Voigt profile components consistent with the errors
\citep{thesis,cvc02}.  In all three systems, the {\MgII} and the
other low and intermediate ionization tranistions could be
fit with one component.  However, the {\CIV} profiles required two
components for the
$z=0.9056$ system, and three for the $z=0.6534$ system.  For selected
transitions, column densities and Doppler parameters of the Voigt
profile fits, performed separately for each transition, are given in
Table~\ref{tab:vptab}.  For the $z=0.6534$ system, numbers are estimated
only for {\MgII} and {\CIV} because fits are ambiguous due to blending.

There is no flux detected from PG~$1634+706$ shortward of a Lyman
limit break at $\sim 1830$~{\AA}, which is due to a strong {\MgII}
absorber at $z=0.9902$ \citep{low1634,ding1634}.  Also, a partial Lyman limit
break at $\sim 1890$~{\AA} reduces the flux to about $30$\% of the
original continuum level due to a multiple cloud weak {\MgII} absorber
at $z=1.0414$ \citep{low1634,zonak1634}.  Although {\OVI} for the $z=0.8181$
system is covered in a $R=10,300$ STIS G230M spectrum of
PG~$1634+706$, this reduced flux and crowding with Lyman series lines
from the $z=0.9902$ system prevent us from deriving a useful
constraint.  We therefore have not used the G230M spectrum as a
constraint in the analysis.


\section{Model Techniques}
\label{sec:techniques}

Several descriptions of the basic technique used for modeling were
given in previous papers \citep{q1206,low1634,weak2}.
For those efforts, only {\MgII}, {\MgI}, and {\FeII} were covered at
high resolution, and all other transitions were observed with FOS/HST
at only $R=1300$.  Even with the availability of numerous transitions
at high resolution, the modeling technique used in the present paper
is very similar to those previous efforts.  We begin with the phase of
gas that produces the dominant {\MgII} absorption, ``the low
ionization phase''.  Then we add in
other phases of gas as needed to reproduce the observed absorption
profiles for all transitions.

For each single--cloud {\MgII} absorber we begin with the column
density, $N({\MgII})$, and the Doppler parameter, $b({\MgII})$ derived
from a Voigt profile fit to the {\MgII} doublet.  Cloudy
photoionization models (version 94.00; \citep{cloudy}) were applied to
determine column densities of the various transitions that would result from
the same phase of gas that produces this {\MgII}, assuming a slab
geometry.  The parameters for
these models are the metallicity $Z$ (expressed in units of the solar
value), the abundance pattern (initially assumed to be solar), and the
ionization parameter, $\log U$.  The ionization parameter is defined
as the ratio of ionizing photons to the number density of hydrogen in
the absorbing gas, $n_H$.  We assume that the ionizing spectrum is of
the form specified by \citet{haardtmadau96}
for $z=1$.  The normalization of the Haardt and Madau background at $z=1$ is
fixed so that $\log n_H = -5.2 - \log U$.
The assumption that stellar sources do not make a
substantial contribution is likely to be valid since weak {\MgII}
absorbers typically do not have a nearby high luminosity galaxy, such
as a starburst.  The effects of alternative spectral shapes are
discussed in \S~\ref{sec:specshape}.

The Doppler parameters of other elements are derived from
$b({\MgII})$, using the temperature output from Cloudy to derive the
thermal and turbulent contributions for each element.  For each choice
of parameters, the Cloudy output column densities and Doppler
parameters are used to synthesize noiseless spectra, convolving with
the instrumental profile characteristic of STIS/E230M.  The synthetic
model spectra are compared with the observed profiles to identify
permitted regions of the parameter space.  The model column densities
and Doppler parameters for permitted models are within $1\sigma$ of
the values measured from the data.

From the three systems studied in this paper, {\FeII} $\lambda2600$ is
covered, but not detected.  For $\log U > -4.0$, the ratio
$N({\MgII})/N({\FeII})$ is strongly dependent on the ionization
parameter in the optically thin regime (see \citet{weak2}) and the
{\FeII} limit can be used to place a lower limit on $\log U$ of
the low ionization phase.  To
obtain an upper limit on the ionization parameter of this phase,
{\SiIV} and {\CIV} provide constraints.  For optically thin gas, the
constraints on $\log U$ do not depend significantly on the
metallicity.  The metallicity is constrained by fitting the {\Lya} and
any higher Lyman series lines that are covered for the system.  Low
metallicities will overproduce the {\Lya} in the wings.  High
metallicities will underproduce the {\Lya}, but cannot be
excluded since the additional {\HI} absorption can arise in a
different phase.  We assumed a solar abundance pattern, but note that
there is a simple tradeoff between metallicity and the abundance of
the metal--line transition that is compared to the hydrogen, discussed
further in \S~\ref{sec:abun}.

In general, once the basic constraints on $\log U$ are derived, we
consider whether a single cloud model can fully reproduce profiles of
all of the observed transitions.  {\it In all three of the systems
modeled in this paper, we will conclude that the {\CIV} absorption
cannot be fully produced in the same phase with the {\MgII}
absorption.}  For the range of permitted values of $\log U$ for the
low--ionization phase, we constrain the properties of the
high--ionization phase that is required to fit the {\CIV} profile.
First, we consider whether a single,
relatively broad component is sufficient.
If the {\CIV} profile is asymmetric (as in the $z=0.9056$ system)
or shows velocity structure (as in the $z=0.6534$ system),
then it is clear that more than
one component is needed.  A Voigt profile fit to the {\CIV} serves as
a starting point for deriving the number of clouds needed in the
high ionization phase, and their
$N({\CIV})$ and $b({\CIV})$.  Results from these fits are listed
in Table~\ref{tab:vptab}.  The $b({\CIV})$ of these {\CIV}
clouds is larger than for the {\MgII} clouds.  Cloudy models of this
additional phase are constrained to match the observed {\CIV} clouds
and $\log U$ is adjusted to determine what range of values are
consistent with other transitions.  An upper limit is set so that
{\NV} and {\OVI} (if covered) are not overproduced.  A lower limit is set so that
the broader components do not produce observable {\MgII} and {\SiII}.
In some cases the intermediate ionization transitions, such as
{\SiIII}, {\CII}, and especially {\SiIV}, could not be fully produced
in the {\MgII} cloud phase.  In these cases, an intermediate value of
$\log U$ is sought for the {\CIV} clouds in order to also account for
the remainder of these transitions, without overproducing the lower
ionization transitions.  A lower limit on the metallicity of a {\CIV} cloud
can be derived in order that {\Lya} absorption is not overproduced.
More exactly, it is the combination
of the {\MgII} and {\CIV} clouds that are constrained not to overproduce
{\Lya} absorption, so there is a trade--off between their metallicity
constraints.
If all transitions of a certain element (such as
{\SiII}, {\SiIII}, and {\SiIV}) are underproduced relative to other
elements, then simple abundance pattern variations
($\alpha$--enhancement and depletion) are considered.

For the high ionization phase, we consider collisional ionization
models as alternatives to Cloudy photoionization solutions.  In this
case, an alternative source of heating (eg., shocks) must be
responsible for heating the gas to the assumed, higher $T$.  The
measured Doppler parameter of the {\CIV}, $b({\CIV})$, is used to
place an upper limit on the temperature, $T < b({\CIV})^2 m({\CIV}) /
2 k$.  For an assumed $T$ and $Z$, and with the measured $N({\CIV})$,
the collisional equilibrium tables of \citet{sutherland93} were used
to determine the column densities of all other covered transitions.
The Doppler parameters of these other transitions, $b_{tr}$ were
calculated from $b_{tr}^2 = 2kT/m_{tr} + b_{turb}^2$, where the
turbulent component of the Doppler parameter is given by $b_{turb}^2 =
b({\CIV})^2 - 2kT/m({\CIV})$.  In order to constrain $T$, synthetic
spectra were generated and superimposed upon the data to facilitate
comparison.  Also, the model column densities of the various
transitions were compared to the measured values.

The {\Lya} profiles provided a constraint on the metallicity of any
collisionally ionized phase.  ``Wings'' on these profiles can be
produced by such a phase, which would be characterized by a relatively
large $b$ parameter.  The contribution of a collisionally ionized
phase to {\Lya} is minimized at low $T$ and high metallicity.  A lower
limit on $T$ is therefore set so as not to overproduce {\Lya} at the
highest reasonable metallicity (usually taken to be solar).
As for photoionization models, for the high ionization phase, the
lower limit on the metallicity depends on the metallicity of
the low ionization phases, which determines its contribution to
{\Lya}.


\section{Model Results}
\label{sec:results}

First, we describe the model results for the two weak, single cloud
{\MgII} absorbers detected in the original HIRES/Keck survey
\citep{weak1}.  Then the results for the newly discovered $z=0.6534$
system are presented.  A range of parameters for satisfactory models
are summarized in Table~\ref{tab:tabmod}.  Model profiles for an example of an
acceptable model are superimposed on the data for each of the three
systems in Figures~\ref{fig:data81}, \ref{fig:data90}, and
\ref{fig:data65}.  Parameters for these sample models are listed in
Table~\ref{tab:tab4}.  Results in this section use the simplest
set of assumptions that produce models consistent with the data,
i.e. Haardt and Madau spectrum and a solar abundance pattern.  The
effects of alternate spectra and abundance patterns are discussed in
\S~\ref{sec:caveats}.

\subsection{The $z=0.8181$ System}
\label{sec:res81}

\begin{figure*}
\figurenum{1} 
\vglue -1in 
\plotone{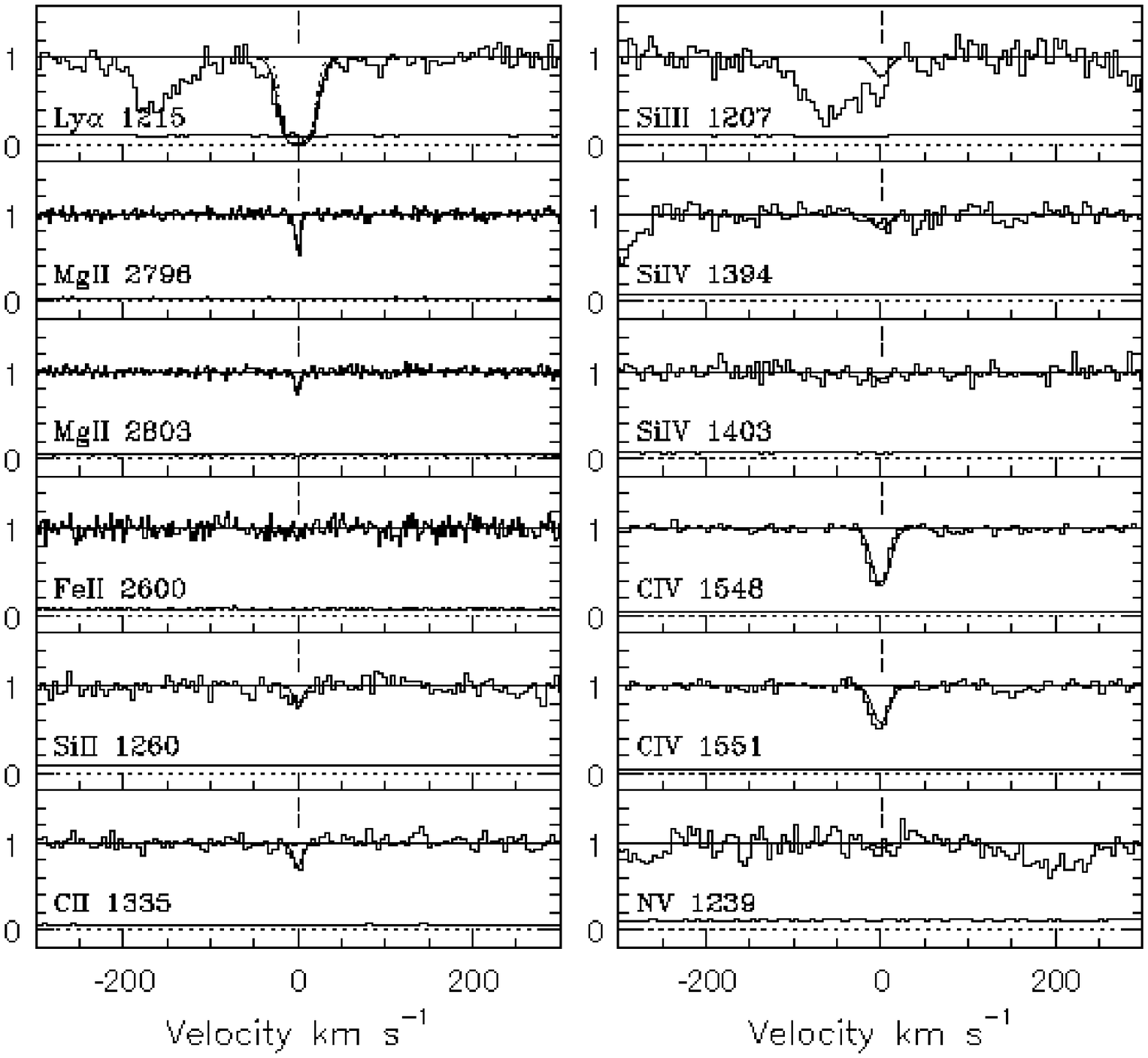} 
\vglue -0.5in
\protect\caption{\small Detected transitions and constraining limits
for the weak, single cloud {\MgII} system with velocity centroid at
$z=0.818157$ are presented in velocity space.  The data are at
$R=45,000$ from HIRES/Keck for the {\MgII}, {\MgI}, and {\FeII}
transitions.  All other transitions are taken from the STIS/HST
spectra, at resolution $R=30,000$.  The sigma spectrum is indicated
as a solid curve just above the dotted line crossing each plot at
zero--flux.  Some transitions that were covered
in the spectrum are not displayed because of severe blending with
stronger transitions.  The position of the lower row of ticks,
displayed above all of the transitions (at zero velocity), was
determined based upon a simultaneous Voigt profile fit to the {\MgII}
doublet.  The upper row of ticks show velocities of the additional
components that were required to fit {\CIV}.
An example of a model (summarized in Table~\ref{tab:tab4}) that provided an adequate fit to the data is
superimposed on the data as a solid curve.  The dotted and dashed curves
represent model contributions of the low ({\MgII}) and high ({\CIV})
ionization phases.
}
\label{fig:data81}
\end{figure*}

\begin{figure*}
\figurenum{3}
\plotone{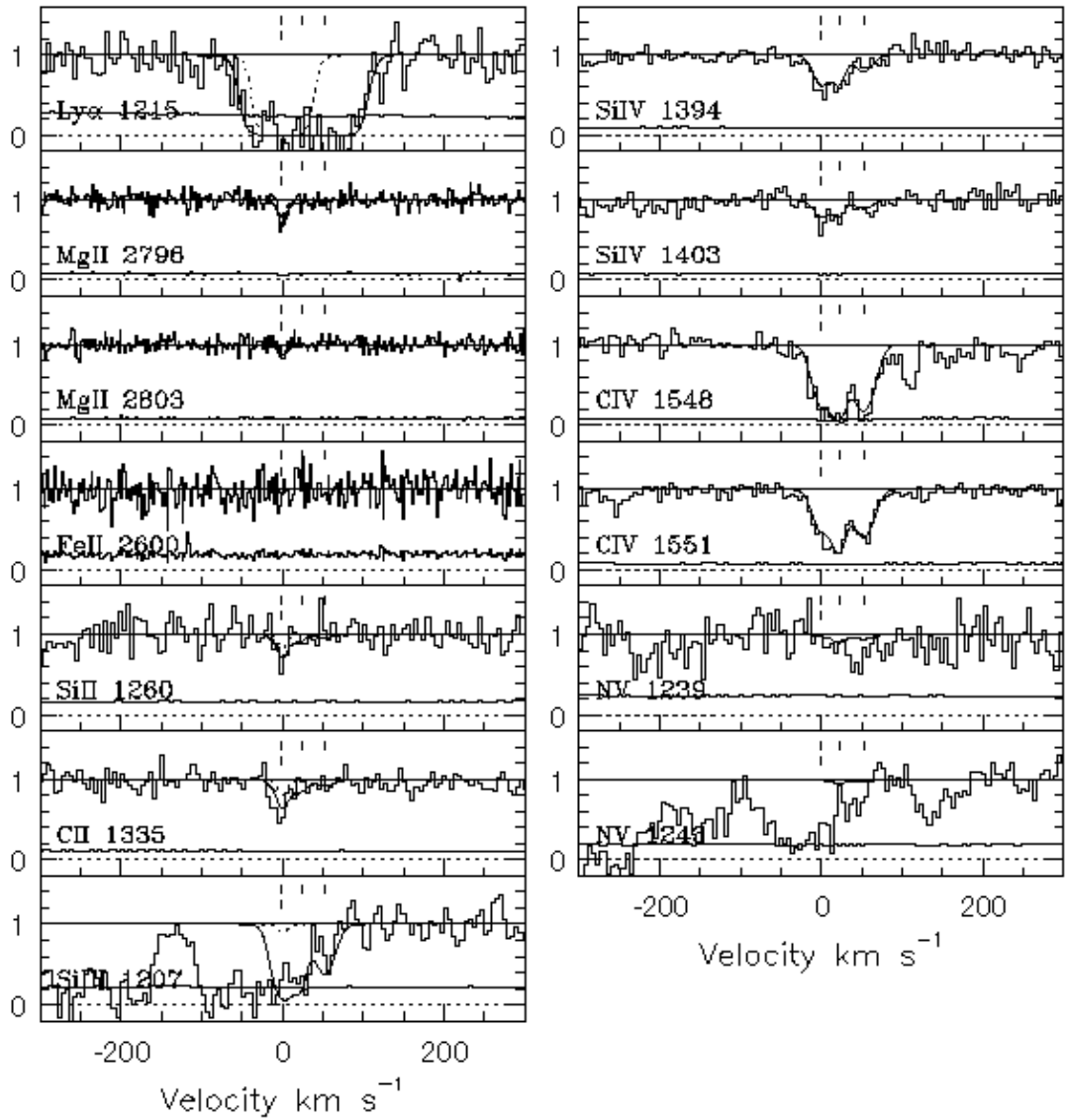}
\vglue -0.4in
\protect\caption{Velocity aligned plot for the system at $z=0.653411$,
displayed as in Figure~\ref{fig:data81}.
}
\label{fig:data65}
\end{figure*}

Figure~\ref{fig:data81} shows detections of {\Lya}, {\MgII}, {\SiII},
{\CII}, {\SiIII}, {\SiIV}, and {\CIV}.  There are useful
limits for {\FeII} and {\NV}.  An obvious, but important, first result is that
{\CIV} is now clearly detected in this system, despite the strong limit
from the earlier low resolution data.

We begin with the Voigt profile fit to the {\MgII} doublet, and adjust
the ionization parameter to fit as many of the other transitions as
possible.  We find that the {\MgII} cloud is constrained to have $\log
U \la -4.0$.  There is no strict lower limit, because the
$N({\MgII})/N({\FeII)}$ vs. $\log U$ curve is flat for $\log U < -4.0$.
Very small values of $\log U$ (as low as $\log U \sim -6.0$) are
permitted, but cloud sizes would be extremely small.  For $\log U <
-6.0$, {\CII} is underproduced.  To find an upper limit on ionization
parameter, we first determined that, for $\log U = -2.5$, the model
would produce minimum fluxes at the positions of {\SiIV} and {\CIV}
that are consistent with the observed profiles.  However, the model
$b({\SiIV})$ and $b({\CIV})$ are small compared to the observed
values, i.e. the model profiles are narrow compared to the observed
profiles.  We therefore conclude that the {\SiIV} and {\CIV} are
produced in a separate higher ionization phase, and that $\log U \la
-4.0$ for the {\MgII} cloud phase.  The contribution of the {\MgII}
cloud phase to the absorption profiles (negligible for all but the singly
ionized transitions) is denoted by a dotted line
in Figure~\ref{fig:data81}.

The metallicity of the cloud with detected {\MgII} is constrained to
be $\sim 0.3$~dex greater than solar (for a solar abundance pattern).
For lower metallicities, absorption in the wings of the {\Lya} profile
will exceed that observed.  Also, unless the metallicity is even
higher for the {\MgII} cloud, the additional high ionization phase
(required to fit {\CIV}) is constrained not to give rise to
significant {\Lya} absorption.  For $\log U = -4.0$ and $\log Z =
0.3$, the cloud size would $\sim 0.1$~pc.  For $\log U = -6.0$ and
$\log Z = 0.3$, the cloud size would be only $\sim 150$~AU, and the
observed {\CII} profile would be somewhat underproduced by the model.

Primarily because of the breadth of the {\CIV} profiles, we concluded
that a second phase is required to fit the {\CIV}, even in this system
for which the {\CIV} absorption is relatively weak.
A Voigt profile fit to the {\CIV} profile yields an adequate fit for
a single cloud with $b({\CIV})\sim10$~{\kms} that is centered on
the {\MgII} cloud.  Details of the fit are listed in Table~\ref{tab:vptab}.
We consider whether the {\CIV} can be produced by photoionization,
and/or whether it can be produced by collisional ionization in gas
that has been heated above the equilibrium value.

For photoionization models of the high ionization phase, we optimized
on the {\CIV} Voigt profile fit values, and the ionization parameter
was constrained by data for other intermediate and high ionization
transitions.  To produce the observed {\SiIV} absorption in this
phase, $-2.2 < \log U < -1.8$ is the optimal range.  The cloud size
would be $\sim100$~pc, considerably larger than the {\MgII} cloud.  If
$\log U < -2.2$, the system would have too much {\SiIII} and {\SiIV}
absorption relative to {\CIV}.  If $\log U > -1.8$, $N({\NV})$
would be overproduced by the model.  The metallicity of the high
ionization phase is constrained, by the {\Lya} profile, to be solar or
higher.  For larger values of the ionization parameter, within the
constrained range of $-2.2 < \log U < -1.8$, the lower limit on
metallicity would be raised to a supersolar value.

Considering collisional ionization, for $b({\CIV}) = 10.4$~{\kms},
pure thermal broadening gives an upper limit on the temperature of
$\log T < 4.90$.  Below this limiting temperature, {\SiIV} would be
severely overproduced \citep{sutherland93} if we optimize on {\CIV}.
However, raising the temperature to $\log T = 4.93$ (which is just
consistent with the Voigt profile fit $b({\CIV})$ within $1\sigma$
errors) reproduces the observed $N({\SiIV})$.  The metallicity of this
collisionally ionized phase would have to be solar or greater in order
that {\Lya} would not be overproduced.  Though it requires
fine--tuning of the temperature, a collisionally ionized phase with
solar metallicity and with $\log T = 4.93$ provides an adequate fit to
the data.

We conclude that the $z=0.8181$ system has two phases, both with a
metallicity solar or higher.  The low--ionization phase, with
$b=2$~{\kms}, has $\log U < -4.0$, while the high--ionization phase,
with $b=10$~{\kms}, could be photoionized with $\log U \sim -2.0$ or
collisionally ionized with $\log T = 4.93$.  The broader
high--ionization phase is centered on the {\MgII} cloud and is
consistent with a single cloud producing the {\CIV} absorption.
Constraints are summarized in Table~\ref{tab:tabmod}.  An acceptable model, including
two photoionized phases, is superimposed on the data
in Figure~\ref{fig:data81}.  Column densities of selected transitions,
produced by this model, are given in Table~\ref{tab:tab4}.

\subsection{The $z=0.9056$ System}
\label{sec:res90}

\begin{figure*}
\figurenum{2}
\plotone{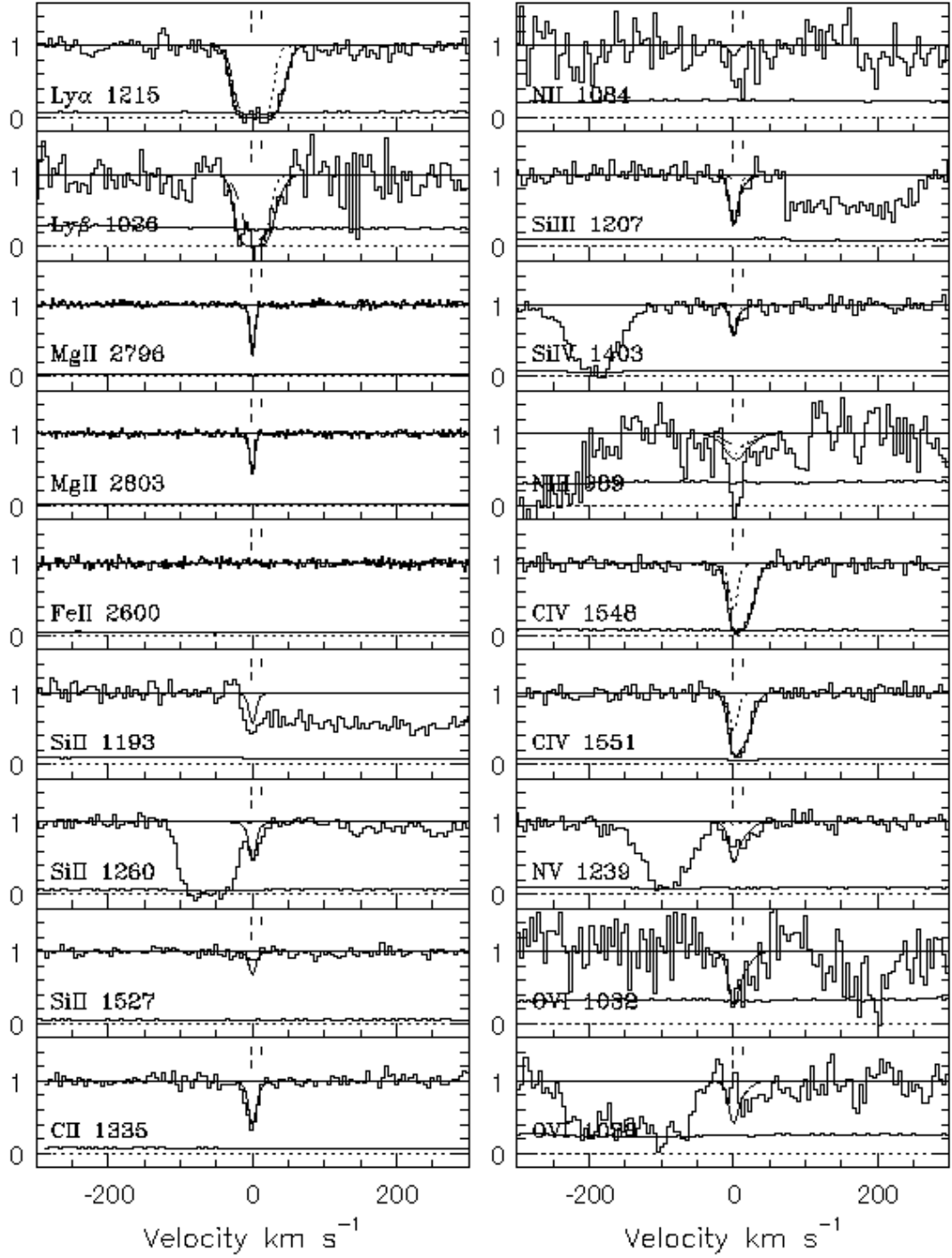}
\vglue -0.2in
\protect\caption{Velocity aligned plot for the system at $z=0.905554$,
displayed as in Figure~\ref{fig:data81}.
}
\label{fig:data90}
\end{figure*}

The $z=0.9056$ system is detected in {\Lya}, {\Lyb}, {\MgII}, {\SiII},
{\CII}, {\NII}, {\SiIII}, {\SiIV}~1403, {\NIII}, and {\CIV}.
{\NV}~1239 is not confirmed by a detection of {\NV}~1242, so it is
viewed as a tentative detection.  {\OVI} is also a likely
detection, although {\OVI}~1038 is in a confused region of the
spectrum that required an uncertain continuum fit.  Limits are
available for {\MgI} and {\FeII}~2600 in the HIRES/Keck spectrum.  All
of these transitions are shown in Figure~\ref{fig:data90}.  In this
system, the {\CIV} is quite strong, and shows an asymmetry in both
members of the doublet.

We first optimize on the $N({\MgII})$ and $b({\MgII})$ given by a Voigt
profile fit to the {\MgIIdblt} profiles (see Table~\ref{tab:vptab}).  Comparing to the other transitions,
the ionization parameter for this {\MgII} cloud is constrained to be
$-3.3 < \log U < -2.7$.  A higher value overproduces {\SiIV}~$1403$
absorption at $v=0$~{\kms}, which is well fit for $\log U$ at the upper end of this
range and underproduced for $\log U < -3.0$. ({\SiIV}~$1394$ is blended
with a stronger transition from another system and cannot be used as a
constraint.)  A value of $\log U < -3.3$ would result in a model that
exceeds the limit on {\FeII}~$2600$.  If $-3.3 < \log U < -3.0$, then
there would have to be a significant contribution to the {\SiIV} absorption from
another phase.  Even for the maximum consistent $\log U = -2.7$,
the {\CIV} absorption is not fully produced in this phase, as shown by
the dotted curve in Figure~\ref{fig:data90}.

The cloud giving rise to the {\MgII}, with $b({\MgII}) = 3$~{\kms},
would overproduce {\Lya} in its blue wing unless it has solar or
super--solar metallicity.  The metallicity constraint could be
slightly relaxed if magnesium is enhanced due to an $\alpha$--enhanced
abundance pattern.  On the other hand, in the red wing, the observed
{\Lya} is not exceeded by any model with $\log Z>-2$.  This difference
between the metallicity constraints from the red and blue wings of the
{\Lya} profile is a consequence of an asymmetry in the distribution of
{\HI} relative to the {\MgII}.  The asymmetry in {\Lya}, along with
the need to fully produce the {\CIV} absorption, requires an additional
phase of gas.

In addition to the {\MgII} cloud, two more clouds are needed in order
to fit the {\CIV} profile.  From a simultaneous Voigt profile fit of
the $1548$~{\AA} and $1550$~{\AA} transitions, the first, centered on
the velocity of the {\MgII} absorption, would have $\log N({\CIV}) \sim
14.0$ and $b({\CIV}) \sim 6$~{\kms}.  A small amount of the {\CIV} absorption is
produced in the same phase with the {\MgII} for $\log U = -2.7$, so the
column density of the {\CIV} cloud would be reduced slightly in that case.
The second {\CIV} cloud, offset by $15$~{\kms} to the red, is fit with
$\log N({\CIV}) \sim 13.9$ and $b({\CIV}) \sim 14$~{\kms}.

First, we consider the $b({\CIV}) = 6$~{\kms} {\CIV} cloud centered on the
{\MgII} profile.  It is too narrow ($\log T < 4.4$) for the {\CIV} to
be produced by collisional ionization.  From Cloudy photoionization
models, to simultaneously fit the {\CIV} and the {\NV} absorption, we
derive the constraint, $-1.8 < \log U < -1.5$, for this high
ionization phase.  If so, little {\SiIV} arises in this phase, which implies
that {\SiIV} absorption should be produced in the low ionization phase
with the {\MgII}.  Revisiting the constraints for the {\MgII} cloud,
we now find the constraint $-3.0 < \log U < -2.7$ for the
low ionization cloud, at the upper end of the previous constrained range
discussed above.  The metallicity of the high ionization gas that produces
the {\CIV} absorption must be high enough not to overproduce {\Lya} in
its blue wing.  In conjunction with a solar metallicity {\MgII} cloud phase,
this yields a solar metallicity for this $-1.8 < \log U < -1.5$ {\CIV} phase
as well.

Next, we consider the $b({\CIV}) = 14$~{\kms} {\CIV} cloud, offset by
$15$~{\kms} from the {\MgII} cloud, that fills in the red wing of the {\CIV}
profile.  
The $b({\CIV})=14$~{\kms} {\CIV} cloud, at $v=15$~{\kms}, can be fit
with a simple, single--cloud photoionization model with  $-1.9
< \log U < -1.8$.  This ionization parameter produces a consistent
fit of the {\OVI}, {\NV}, and {\CIV}, without overproducing the
{\SiIV}.  The {\SiIII} and {\SiIV} data are very slightly underproduced
by the model at this velocity, suggesting that a small ($\sim 0.2$~dex) abundance
pattern enhancement of silicon might be needed.
A simple model with the minimum number of phases
would call for the asymmetry in the {\Lya} profile to be produced
in this cloud along with the red wing of the {\CIV}.
The metallicity is thus constrained to be at the solar
value in order to fit the red wing of the {\Lya} line.

Our philosophy of fitting with the minimum number of phases argues against
collisional ionization as the mechanism for producing the observed
{\CIV} absorption at $v \sim 15$~{\kms}.  To be consistent with this
philosophy, both {\CIV} and {\Lya} should arise in the same phase.
The limit on $\log T$ for this
component with $b({\CIV})=14$~{\kms} is consistent with production of {\CIV}
absorption by collisional ionization.  However, for the
maximum permitted $T$ there would not be enough {\HI} absorption in
this model component unless the metallicity was well over supersolar.  With the
relatively low $T$ needed to produce the much {\HI} absorption,
{\SiIV} absorption would be overproduced.

We conclude that the $z=0.9056$ system must have solar
metallicity or higher in its low--ionization phase.  The $b({\MgII})=3$~{\kms}
cloud that produces {\MgII} absorption must be of a relatively high
ionization state, with $-3.0 < \log U < -2.7$.  The range of derived
cloud sizes for this constrained range is $30$--$100$~pc.  Two higher
ionization, broader clouds are also required: one, with
$b({\CIV})\sim6$~{\kms}, is centered on the {\MgII} cloud, while the other
$b({\CIV})\sim14$~{\kms} cloud is offset by $15$~{\kms}.  Both are consistent
with photoionization with $\log U \sim -1.8$, and have sizes of $0.4$--$1$~kpc.
Ranges of acceptable model parameters are presented in Table~\ref{tab:tabmod} and model column
densities contributed by the three clouds for an adequate model are listed in Table~\ref{tab:tab4}.
The predictions for this same model are superimposed on the data in
Figure~\ref{fig:data90}.

\subsection{The $z=0.6534$ System}
\label{sec:res65}

The {\CIV} profile for this newly discovered system has a
complex structure and the {\MgII} is extremely weak and narrow.
The {\CIV} profile requires at least a three component fit.
The {\Lya} is not symmetric about the {\MgII}, but it is approximately
symmetric about the {\CIV}.  These profiles as well as those of
the other detected transitions, {\SiII}, {\CII}, {\SiIV}, are
displayed in Figure~\ref{fig:data65}, along with the region of
the spectrum that provides a limit on {\FeII}~2600.
{\NV} is in a noisy region of the spectrum.  If the detected
feature is really {\NV}, it is present only in the redward component,
at $\sim50$~{\kms}.

We begin by considering the {\MgII} cloud.  
Since the {\FeII}~2600 spectrum is noisy, the low ionization parameter
constraint on $\log U$ is not strong; for $\log U < -4$,
$N({\MgII})/N({\FeII})$ is relatively constant.  However, assuming
solar abundance pattern, $\log U > -4$ provides a marginally better
fit to the {\SiII} and {\CII}.  If $\log U \sim -2.5$, then the depth
of the {\SiIV} and {\CIV} absorption at $0$~{\kms} can be reproduced
in the {\MgII} cloud, but the model profile is too narrow and does not
provide a good fit to the data.  A separate higher ionization phase is
needed to fit these higher ionization transitions.  Considering all
transitions, and assuming solar abundance pattern, $-4 < \log U <
-3$ is favored for the {\MgII} cloud.

The metallicity of the {\MgII} cloud would be $\log Z = -1.5$ if all
of the {\Lya} in the red wing were to arise in this cloud.  However,
it could also be significantly higher if there were {\Lya}
contribution from the {\CIV} clouds.  For $\log Z = -1.5$ and $-4 < \log U
< -3$, the {\MgII} cloud size ranges from $6$~pc to $\sim 1$~kpc.  Sizes scale
with $Z^{-1}$ so that $\log Z = -1.0$ clouds would be a factor of
$\sim3$ smaller than $\log Z = -1.5$ clouds.

Consider the case of $\log U = -4.0$ for the {\MgII} cloud.  Three
more clouds are required to fit {\CIV}.  The column densities and
Doppler parameters for these three clouds were obtained with a Voigt
profile fit. This fit is not unique because clouds cannot be well
separated due to blending, so there are no errorbars listed in Table~\ref{tab:vptab}.
The cloud centered on the {\MgII} was fit with $\log N({\CIV}) =
13.7$, $b ({\CIV}) =13$~{\kms}.  The other two clouds, at $v=24$~{\kms}
and $v=54$~{\kms}, were fit with $\log N({\CIV}) = 13.9$,
$b({\CIV})=9$~{\kms}, and $\log N({\CIV}) = 13.8$, $b({\CIV}) =
14$~{\kms}, respectively.

For photoionization models, the ionization parameters for these three
{\CIV} clouds are constrained by the requirement that they produce the
observed {\SiIV} without producing significant lower ionization
transitions.  The $b({\CIV}) =13$~{\kms} {\CIV} cloud, centered on the
{\MgII} cloud, is tightly constrained to have $-2.5 < \log U < -2.4$,
by the {\SiIV} profiles.  The range already takes into account the
fact that the {\MgII} cloud would make a small contribution to {\SiIV}
if its $\log U = -3$.  The lower limit on $\log U$ of the {\CIV} cloud
arises in order that the {\MgII} is not overproduced in its wings by
this relatively broad component.  The upper limit applies in order to
produce sufficient {\SiIV} absorption.

The $b({\CIV})=9$~{\kms} {\CIV} cloud at $v=24$~{\kms}, has $-2.2 < \log U <
-2.1$, in order to produce the optimal fit to {\SiIV}, {\CII}, and
{\CIV}.  For $\log U > -1.7$, {\NV} is overproduced, and for $\log U <
-2.6$, {\CII}, {\SiII}, and {\MgII} are overproduced.

For the $b({\CIV})=14$~{\kms} {\CIV} cloud at $v=54$~{\kms}, $\log U =
-2.2$ provides an optimal fit to the {\SiIV}, and is consistent with
{\CII} and with the limit on {\NV}.  A value of $\log U = -2.0$ is
only marginally consistent, with the model slightly underproducing the
{\SiIV}.  This inferred properties of this cloud are
independent of the {\MgII} cloud since no {\MgII} is detected at this
velocity.

As with the {\MgII} cloud, the metallicities of the three additional
{\CIV} clouds cannot be extremely low.  Only $\log Z = -1$ or higher is
consistent with the data.  If the red side of the {\Lya} is to be
produced by the $v=54$~{\kms} {\CIV} cloud, $\log Z = -1.0$ would
apply for this cloud.  Similarly, $\log Z = -1.0$ for the $v=0$~{\kms}
{\CIV} cloud would match the blue wing of {\Lya}.  In the case of the
blue wing, the {\Lya} could in principle be produced in a $\log Z =
-1.5$ {\MgII} cloud and the metallicity of the {\CIV} cloud could be
higher.  Constraints on the metallicities of these clouds, relative to
the cloud that produced the narrower {\MgII} profile, are limited by
the lack of coverage of higher order Lyman series lines.  However, the
data are consistent with $\log Z = -1$ in the {\MgII} cloud and in the
three {\CIV} clouds.  For this metallicity, the sizes of the three
{\CIV} clouds would be $\sim2$--$4$~kpc.

We also consider whether collisionally ionized gas can be consistent
with the observed profiles of {\CIV}, {\SiIV}, and {\Lya}.
For the $v=0$~{\kms} {\CIV} cloud, $\log T \ge 4.90$ does not overproduce
{\SiIV}, and $\log T = 4.90$ is consistent with producing the blue side of
the {\Lya} profile for $\log Z = -1$.  If the metallicity is higher,
a higher ionization parameter is needed to fit the {\Lya} line, but in
this case the model {\Lya} profile shape is not consistent with the data.
The $v=24$~{\kms} cloud is too narrow to be consistent with production
of {\CIV} absorption through collisional ionization assuming the
particular Voigt profile fit that we have adopted.  However, because
this fit is not unique, there could possibly be a broader cloud component
that could be reconciled with collisional ionization.  Finally, the
$v=54$~{\kms} {\CIV} cloud could be collisionally ionized with
$\log T = 4.95$ and $\log Z = -1$, in which case it would match
the red side of the {\Lya} profile.

In summary, the $z=0.6534$ system {\MgII} profile is quite weak and narrow.
However, the {\CIV} profile can be fit with several
components spread in velocity over $50$~{\kms}.  The three
components are similar to each other, having $b\sim 10$~{\kms} and
$\log U$ ranging from $-2.5$ to $-2.1$ if they are photoionized.
Collisional ionization with temperatures of $\log T \sim 4.9$ could be
consistent with the data, but requires fine--tuning of the temperature
and seems less likely.  The bluest {\CIV} component has the {\MgII}
detected at the same velocity, but in a narrower component
($b\sim4$~{\kms}) produced in gas with a somewhat lower ionization parameter
($\log U \sim -3.5$).  The {\CIV} clouds have a metallicity of at
least $\log Z = -1$ and must be at least a couple of kiloparsecs in
size.


\section{Caveats}
\label{sec:caveats}

\subsection{Abundance Pattern Variations}
\label{sec:abun}

Abundance ratios measured in Galactic stars show a clear range of 
$0 \leq [\alpha/{\rm Fe}] \leq +0.5$ \citep{jtl96}.
In our presentation of model results (\S~\ref{sec:results})
solar abundance pattern was assumed unless the data require an
alternative pattern.  However, it should be noted that the inferred
metallicity depends directly on the assumed abundance pattern.
If, instead of solar, the abundance pattern is $\alpha$--group enhanced
with $\hbox{[$\alpha$/Fe]} = +0.5$, the inferred metallicity of a
phase constrained by $N({\MgII})$ (an $\alpha$--group element)
would proportionally drop by $\sim 0.5$ dex.
Constraints on the ionization parameter, $\log U$, are also affected
by changes in the assumed abundance pattern.  For example, if the
the abundance pattern is $\alpha$--group enhanced, the
ratio $N({\SiIV})/N({\CIV})$ would constrain $\log U$ to be
larger than if the abundance pattern is solar.  However, for
$\hbox{[$\alpha$/Fe]} \le +0.5$, the change in the constraint
on $\log U$ (and therefore on $\log n_e$) is less than $0.3$ dex.
We have used the relative absorption strengths in several other
transitions to determine $\log U$, but $0.3$ dex is typical of
the level of uncertainty due to variations in abundance pattern.

\subsection{Effects of Spectral Shape}
\label{sec:specshape}

Based on results for other similar absorption line systems, there is
likely to be no bright galaxy associated with these single--cloud weak {\MgII}
absorbers \citep{weak2}.  Therefore, we used the \citet{haardtmadau96} extragalactic
background spectrum for $z=1$ in
our detailed models presented above (\S~\ref{sec:results}).
However, because it is not strictly excluded, we do consider
here the effect of changing the spectral shape.

Two alternative galaxy spectra, $0.01$~Gyr and $0.1$~Gyr instantaneous
starburst models, were superimposed on the Haardt and Madau
extragalactic background spectrum.  Both models, with solar
metallicity and a Salpeter IMF were taken from \citet{bc93}.
The normalizations of the starburst spectra are
defined relative to the Haardt and Madau spectrum at $1$ Rydberg, and
in all cases the extragalactic and galactic contributions were
superimposed.  The largest reasonable value for the photon flux from
even the most extreme starburst galaxy is $10^{54} {\rm s}^{-1}$, and
of order $1$\% of the photons escape \citep{starburst}.  Using these
numbers, to have a flux ten times that of Haardt and Madau at $z=1$,
the absorber would have to be within $6$~kpc of the starburst.  Within
this distance, we would expect stronger absorption than observed in
these weak {\MgII} absorbers.  Therefore, we consider it most likely
that the Haardt and Madau spectral shape is the appropriate
assumption.  Nonetheless, in this section, we briefly examine the
consequences of the alternatives.  We consider the effect of spectral
shape on the metallicity of the low ionization phase, on whether the
{\CIV} absorption can arise from the same phase as the {\MgII}
absorption, and on the ionization conditions of the high ionization
phase.

The largest impact on the metallicity would be from an ionizing
spectrum with a large feature at the Lyman edge, such as the
Bruzual and Charlot $0.1$~Gyr instantaneous burst model.
In this case, the cloud would be more neutral so that not as
much Hydrogen would be required to fit the {\Lya} profile.
In such a case, the metallicity would be even higher than we
inferred assuming a pure Haardt and Madau spectrum.  In general,
we found that the starburst spectrum normalization needed to be
at least ten times that of the Haardt and Madau spectrum in order
to detect a difference in the models.  Even with a normalization
of twenty--five the difference in the inferred metallicity is
less than $0.5$ dex.  Most importantly, even if the spectrum of ionizing
radiation has a substantial Lyman edge, the metallicity of the weak
{\MgII} clouds would be even higher than the solar value --- an
even more surprising result.

A $0.1$~Gyr instantaneous burst model, if it dominates over the
Haardt and Madau spectrum by a factor of twenty--five, can make
significantly more {\CIV} relative to {\MgII}.  However, only
the column density can be made consistent with the observed {\CIV}
profiles in the three systems studied here.  The Doppler parameters
of these lines are still too large for
them to be produced in the same phase with the {\MgII}.

A spectrum with a sharp {\HeI} edge will have the largest
effect on the inferred $\log U$ of a photoionized high ionization phase.  
A $0.01$~Gyr starburst model is extreme in this respect, yet it takes a
normalization of ten times Haardt and Madau to see even a small
change.  With a normalization of twenty--five, the {\SiIV} is
significantly overproduced relative to {\CIV} for the same
$\log U$.  However, the qualitative result of a relatively
low density phase producing the bulk of the {\CIV} is unchanged.
For example, for the $z=0.8181$ system, $\log U$ must be increased
from $-1.5$ to $-1.0$ in order that {\SiIV} is not overproduced.

We conclude, that at $z=1$ an extreme starburst spectrum would have to
dominate in order to affect the conclusions of our models.  Even if
such conditions prevailed, the conclusions would be qualitatively
unchanged, only changing constraints by $\sim0.5$ dex in parameter
spaces ranging over a few orders of magnitude.


\section{Discussion}
\label{sec:discussion}

\begin{figure*}
\figurenum{4}
\plotone{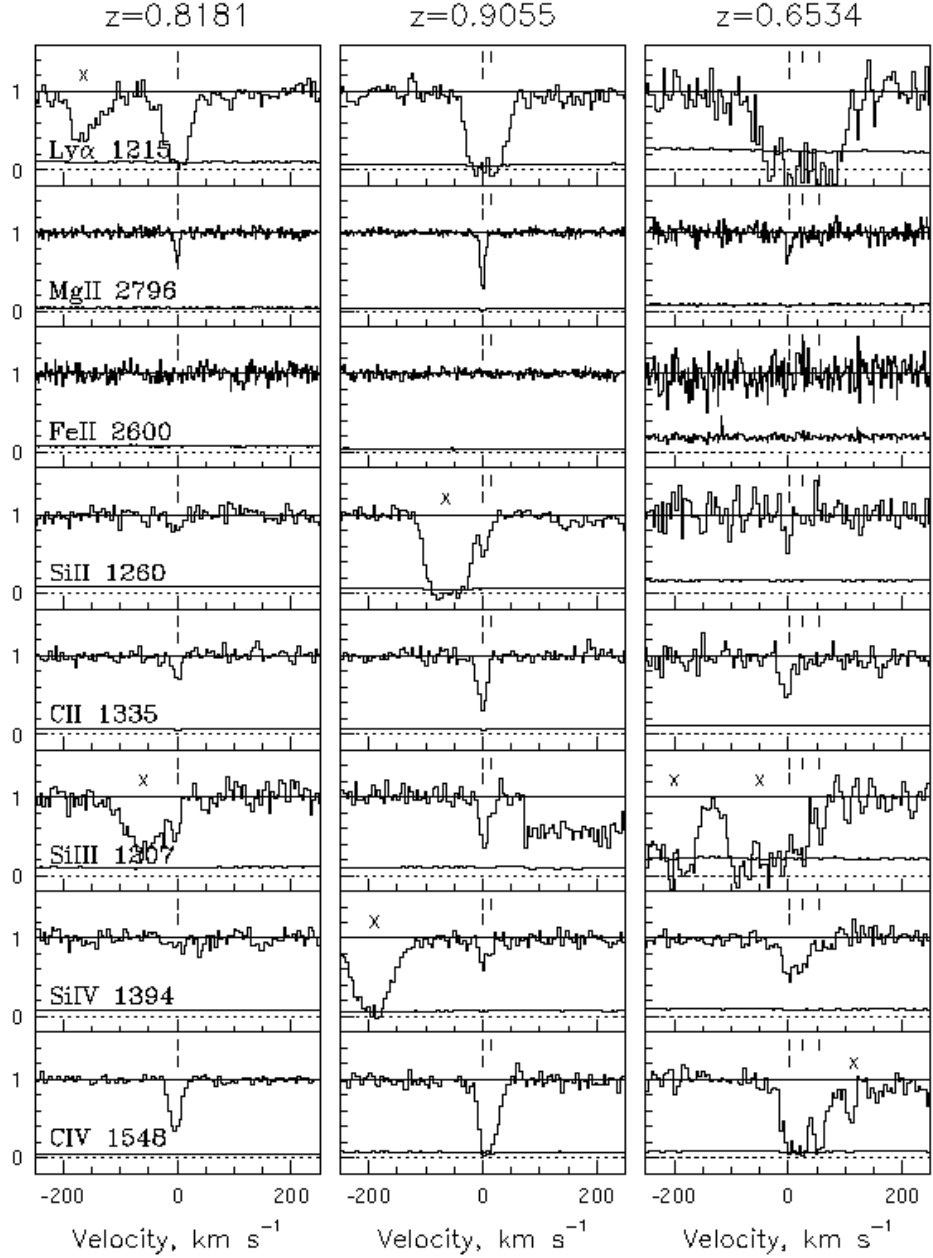}
\vglue -0.4in
\protect\caption{Comparison plot of key transitions for the three weak,
single cloud {\MgII} absorbers along the PG~$1634+706$ line of sight.
The symbol ``X'' denotes a known blend.
Figures~\ref{fig:data81}--\ref{fig:data65}
give more complete separate compilations for each of the systems.
}
\label{fig:allsys}
\end{figure*}

This study of the multiple phases of gas and the physical conditions
of the three weak {\MgII} systems along the PG~$1634+706$ line of
sight indicates a heterogeneous population of objects selected by a
weak {\MgII} doublet.  Figure~\ref{fig:allsys} is a comparison between
the three systems showing the range of {\CIV} equivalent widths and
kinematic structures, ranging from a profile consistent with a single
cloud for the $z=0.8181$ system, to the stronger asymmetric profile
for the $z=0.9056$ system, to the multiple component profile for the
$z=0.6534$ system.  The {\Lya} profile strength does not
systematically increase in proportion with the {\MgII} absorption.
It could, however, be increasing with the kinematic spread of
the {\CIV}. 
Table~\ref{tab:tabmod} gives a summary of the range of
``acceptable models'' for each of the three systems, while
Table~\ref{tab:tab4} gives more detailed information about a sample model that is
within the acceptable range.  That typical model was also superimposed
on the data in Figures~\ref{fig:data81}--\ref{fig:data65}.

\subsection{Comparisons of the Three Systems}
\label{sec:comparisons}

The $z=0.8181$ system and the $z=0.9056$ system provide contrasting
examples of a weak {\MgII} absorber that clearly requires a separate
broad phase to fit a strong {\CIV} profile and one with much weaker
{\CIV}, not even detected in the previous low resolution FOS spectrum
\citep{archive1}.  For both systems, in the present study we find
that a separate broader phase ($b\sim10$~{\kms} and $6$~{\kms}), with
an ionization parameter $\log U \sim -2$, is needed to fit the
observed {\CIV} STIS profile.  This phase is centered at the same
velocity as the phase that gives rise to the weak {\MgII} absorption.
However, in the $z=0.9056$ system there is an asymmetry to the {\CIV}
profile that indicates an additional offset component, $13$~{\kms}
redward of the {\MgII} cloud and the first {\CIV} broad phase.  The
{\CIV} profile can be fit with a two cloud model, though it is also
possible that there is a more complex distribution of material, spread
over a range of velocities, that gives rise to such an asymmetric
profile shape.  A general point is suggested by the comparison of the
$z=0.8181$ and $z=0.9056$ systems.  The strong {\CIV} absorption that
is observed in many weak {\MgII} absorbers \citep{weak2} could be due
to the presence of separate clouds that are of higher ionization and
do not give rise to detectable {\MgII} absorption.  The same would
apply for the stronger {\Lya} absorption apparent in some weak {\MgII}
absorbers.

The $z=0.6534$ system is a more extreme example in which three
separate components are clearly apparent in the {\CIV} profile.  A
narrow, weak {\MgII} absorber is centered on one of the higher
ionization {\CIV} clouds.  The separate {\CIV} components at
$24$~{\kms} and $54$~{\kms} are quite distinct from the {\CIV}
centered on the {\MgII} cloud.  Like the higher ionization components
of the other two systems, these two offset clouds have ionization
parameters, $\log U \sim -2$.  The {\Lya} profile is consistent with
being centered around the three {\CIV} component, but not around the {\MgII}
cloud.  The fact that {\Lya} absorption is strongest in this system
suggests that the spread in velocity of absorbing gas along the line
of sight is an important factor in determining the equivalent width of
{\Lya} absorption.

The metallicities of the $z=0.8181$ and $z=0.9056$ systems are
constrained to be at least as high as solar.  The $z=0.6534$ system is
likely to have a metallicity of at least $1/10$th solar.  These
relatively high metallicities
appear to be common for weak {\MgII} absorbers.  Rigby {\etal} (2001)
inferred a high metallicity for several systems based upon low
resolution FOS data, and in no case did they find that a metallicity
less than $1/10$th solar was required.

\subsection{Comparison to Model Results Based on Low Resolution Data}
\label{sec:confrontlow}

For the $z=0.8181$ system and the $z=0.9056$ system, we can compare
the results from this study, incorporating high resolution STIS data,
to our previous models, based upon the lower resolution FOS data
\citep{low1634}.

For the $z=0.8181$ system, {\Lya} was not covered in the earlier FOS
spectrum, so no constraints on metallicity were available.  The {\CIV}
was not detected in the FOS spectrum, but we showed that it might
be detected in the higher resolution STIS spectrum either from a
{\MgII} cloud with a high ionization parameter or from a broader,
separate phase.  No specific predictions could be made as to which
possibility was more likely.  Now, with the high resolution STIS
spectrum coverage of {\CIV} we are able to place specific constraints
as was outlined in \S~\ref{sec:res81}.

For the $z=0.9056$ system the main conclusions of our previous study,
based on the FOS spectrum \citep{low1634}, were that the {\CIV} is
present in a separate broader phase, and that the metallicity of the
{\MgII} cloud is solar or higher.  These conclusions are confirmed by
the present study.  The detailed properties of the broader phase were
more difficult to determine based upon low resolution spectra.  We
suggested that the broad phase has $b\sim20$~{\kms} because the
doublet ratio of the {\CIV} was large compared to observations, if
$b=10$~{\kms} was assumed.  The high resolution STIS spectrum shows
that the {\CIV} profile is asymmetric, so that it must be composed of
at least two separate ``clouds'', the broadest having
$b\sim14$~{\kms}.  The ``wings'' of the {\Lya} profile in the low
resolution spectrum were apparently due to FOS fixed pattern noise
rather than to an extremely broad phase, since these features are not
apparent in the high resolution STIS spectrum.

From this very limited comparison we tentatively conclude that it
should be possible to draw inferences about the presence of a separate
{\CIV} phase based on a combination of high and low resolution
spectra, i.e. drawing on the large existing FOS database.  Also, our
conclusion of solar metallicity for the $z=0.9056$ system was
confirmed by modeling of the high resolution STIS profile of {\Lya}.
This is important, since modeling of FOS data was used to infer that
many weak {\MgII} absorbers have close to solar metallicity \citep{weak2}.

\subsection{Relationships Between Phases}
\label{sec:naturephases}

To understand the nature of these systems, we seek to infer the spatial
distribution of absorbing material, and the relationship to 
star--forming objects.  All three single--cloud weak {\MgII} absorbers
have two phases that produce a narrower ($2$--$4$~{\kms}) and a broader
($6$--$13$~{\kms}) absorption component centered at the same velocity.
The narrower component is of lower ionization than the broader
component.  For a fixed Haardt--Madau spectrum intensity, this implies
a higher density ($\sim 0.01$--$0.1$~{\cc}) for the narrower component, and
also a smaller size, $\sim 0.1$--$100$~pc.  The broader component would
arise in a higher ionization/lower density phase ($\sim 0.001$~{\cc})
with a larger size ($\sim 0.1$--$2$~kpc).  

Two simple scenarios could be consistent with the inferred properties
of the narrow and broad components: 1) The first would have the lower
ionization region embedded within the higher ionization region, and
the higher ionization region would present a larger cross section.  We
would then expect many systems to be observed for lines of sight that
pass through only the higher ionization region of such structures.
These ``{\CIV}--only systems'' might typically have lower {\CIV}
column densities than two--phase weak {\MgII} systems. They would be
produced at large impact parameter in the structure, at which the
pathlength would be shorter and the gas densities would be lower.  2)
In the second scenario the lower ionization components could be
produced by parts of a shell structure surrounding a lower density,
higher ionization region.  The covering factors for the low ionization
shell fragments would be limited ($<<1$) by the lack of observation of
many two--cloud weak {\MgII} systems.  However, with such a small
covering factor, we would again expect a large incidence of
``{\CIV}--only systems'' which in this scenario would tend to have
similar {\CIV} phases to those of two--phase weak {\MgII} systems.

We searched the PG~$1634+706$ spectra and found no ``{\CIV}--only systems''
with {\CIV} comparable in strength even to that of the $z=0.8181$ weak
{\MgII} absorber.  However, we have identified several weaker candidate {\CIV}
doublets at $>5 \sigma$, confirmed by a line detected at the
expected position of {\Lya}.  Many other lines of sight need to be
systematically surveyed, but there clearly will be limits on the
geometry and covering factors of the two phases of gas.

An alternative to the idea of embedded phases is to have separate
clouds along the line of sight with different densities and sizes.
This is consistent with the presence of an offset cloud in the
$z=0.9056$ system, and the spread of three clouds over $50$~{\kms} in the
$z=0.6534$ system.  These separate clouds could exist as condensations
in larger structures with velocity dispersions of tens of {\kms},
e.g., dwarf galaxies.  However, if the phases were all completely
separate from each other then it would be hard to explain the close
alignment in velocity of the lower ionization cloud with one of the
higher ionization clouds.

The metallicities of these absorbing structures present a clue as to
their place of origin.  We do not expect that they are in the vicinity
of luminous galaxies ($>0.05L^{*}$ galaxies are not found within
impact parameters of $\simeq 50 h^{-1}$~kpc from the quasar).
Although no useful image is available for the PG~$1634+706$ field in
particular, other single--cloud weak {\MgII} absorbers are rarely
found near such luminous galaxies
(C. Steidel, private communication; \citet{weak2}).  Dwarf
galaxies have metallicities significantly lower than solar, which
would appear inconsistent with solar metallicities for single--cloud weak
{\MgII} absorbers.  However, it is possible that the weak {\MgII} absorbers
are concentrations of higher metallicity within lower metallicity
structures.

Most strong {\MgII} absorbers ($W({\MgII}) > 0.3$~{\AA}) also require
a phase in addition to the {\MgII} clouds in order to produce the
observed {\CIV} absorption.  These absorbers are associated with
$\sim L^{*}$ galaxies \citep{bb91,sdp94,s95},
and the high--ionization phase is inferred to have
an ``effective Doppler parameter'' or velocity spread of $\sim50$~{\kms}
\citep{archive2}.  At high resolution some of the {\CIV}
profiles will separate into multiple components, while others may be
due to a more uniform distribution of gas \citep{ding1206,ding1634,zonak1634}.
The {\CIV} phase of some strong {\MgII} absorbers is reminiscent of what would be
expected for a corona such as that observed in {\CIV} and {\OVI}
absorption around the Milky Way disk.  The single cloud weak
{\MgII} absorbers, although they do have a second phase, do not
appear to be related to such a corona.


\section{Conclusions}
\label{sec:conclude}

The combination of STIS/HST and HIRES/Keck high resolution spectra,
covering multiple chemical transitions, provided the first opportunity
to collect direct information on the metallicities and phase
structure of weak {\MgII} absorbers.  There are three weak {\MgII}
absorbers, at $z=0.8181$, $z=0.9056$, and $z=0.6534$ along the
line of sight toward the quasar PG~$1634+706$.
All three of these absorbers
have a second, higher ionization phase, giving rise to the {\CIV}
absorption.  The broad phase in one case is consistent with a
single cloud, and in the other two cases requires one or two additional
clouds separated in velocity space from the one aligned with
the {\MgII} absorption.

Two of the weak {\MgII} absorbers are
constrained to have solar or greater than solar metallicity, and the other one to have
a metallicity greater than $1/30$th solar. Thus, in general, weak {\MgII}
absorbers are not weak because of a low metallicity (confirming
the result of \citet{weak2}).
As introduced in \S~\ref{sec:intro}, it is also possible
that they have weak {\MgII} absorption because they are
more highly ionized than their strong counterparts, or
because their total column densities are smaller.
The likely answer is that there is some combination of
these two effects, perhaps leading to different populations
of weak absorbers arising in different environments.
The ionization parameters of the absorbers studied here
are higher than inferred for many of the clouds in
strong {\MgII} absorbers for which {\FeII} is also
detected.  [{\FeII} is a tracer of low ionization
conditions \citep{weak1,weak2}.]  However, there
is also a population of weak
{\MgII} absorbers with strong {\FeII} lines, amounting
to about one--third of the weak {\MgII} absorber population
\citep{weak2}.  Detailed study of the phase structure
of this sub--group awaits spectra of additional quasar
lines of sight with {\CIV}, {\Lya} and other transitions
covered at high resolution.

Weak {\MgII} absorbers are potentially of general importance
because they provide a sensitive probe of particular types
of star forming environments.  In principle, this population
of objects can be used to track the chemical and ionization
history of the universe in regions that are not in luminous
galaxies which can be studied by other methods.  
For example, do they exist with solar metallicity to high
redshifts, i.e. are there selected environments with
extreme enrichment even at early times?
Answering this question will require a large systematic study of many
weak {\MgII} systems over a range of redshifts.  Such a
study could address whether there is always {\CIV} absorption
centered in velocity on a {\MgII} cloud.  It could tabulate
the distribution of {\CIV} clouds in velocity space, and
address whether a large {\CIV} equivalent width is typically
due to the presence of multiple clouds along the line of
sight.  Spectral coverage of {\NV} and {\OVI} will also provide better
constraints on the ionization conditions of the high--ionization
phase.  Finally, it is highly desirable to search the quasar
fields for faint galaxies that could be related to these
weak {\MgII} absorbers.

\acknowledgements
Support for this work was provided by the NSF (AST--9617185) and by NASA
(NAG 5--6399 and HST--GO--08672.01--A), the latter from the
Space Telescope Science Institute,
which is operated by AURA, Inc., under NASA contract NAS5--26555.
S. Zonak, J. Rigby, and N. Bond were supported by an NSF REU Supplement.
We thank C. Howk and K. Sembach for their invaluable guidance on the analysis
of high resolution STIS spectra.



\clearpage

\clearpage

\clearpage


\newpage
\begin{deluxetable}{lccc}
\tablenum{1}

\tabletypesize{\footnotesize}
\tablewidth{0pt}
\tablecaption{Rest Frame Equivalent Widths [{\AA}]}
\tablehead{
\colhead{Transition} &
\colhead{$z=0.8181$} & 
\colhead{$z=0.9056$} &
\colhead{$z=0.6534$} \\
}

\startdata
\hline
{\Lya}        & $0.220\pm0.010$ & $0.349\pm0.008$  & $0.665\pm0.022$ \\
{\Lyb}        & \nodata         & $0.206\pm0.022$  & \nodata         \\
{\MgI}~2853   & $<0.004$        & $<0.002$         & $<0.010$        \\
{\MgII}~2796  & $0.030\pm0.005$ & $0.034\pm0.002$  & $0.031\pm0.006$ \\
{\MgII}~2803  & $0.018\pm0.006$ & $0.024\pm0.003$  & $0.007\pm0.003$ \\
{\FeII}~2600  & $<0.008$        & $<0.003$         & $<0.022$        \\
{\SiII}~1193  & \nodata         & $0.015\pm0.005$ & $<0.028$        \\
{\SiII}~1260  & $0.020\pm0.004$ & $<0.061$         & $0.021\pm0.007$ \\
{\CII}~1335   & $0.017\pm0.002$ & $0.31\pm0.01$  & $0.045\pm0.005$ \\
{\NII}~1084   & $<0.028$        & $0.063\pm0.016$  & \nodata         \\
{\SiIII}~1207 & $0.045\pm0.007$ & $0.052\pm0.006$  & \nodata         \\
{\SiIV}~1394  & $0.019\pm0.003$ & \nodata          & $0.123\pm0.006$ \\
{\SiIV}~1403  & $0.019\pm0.003$ & $0.026\pm0.003$  & $0.071\pm0.006$ \\
{\NIII}~989   & \nodata         & $0.084\pm0.018$  & \nodata         \\
{\CIV}~1548   & $0.079\pm0.003$ & $0.196\pm0.007$  & $0.355\pm0.025$ \\
{\CIV}~1551   & $0.060\pm0.003$ & $0.154\pm0.004$  & $0.305\pm0.008$ \\
{\NV}~1239    & $<0.004$        & $0.061\pm0.004$  & $0.048\pm0.014$ \\
{\NV}~1243    & \nodata         & $0.027\pm0.006$  & $<0.054$        \\
{\OVI}~1032   & \nodata         & $<0.076$         & \nodata         \\
\hline
\hline
\enddata
\vglue -0.05in

\tablecomments{
\baselineskip=0.7\baselineskip
Limits are at a $3\sigma$ level. The symbol {\nodata}
indicates that the transition was not covered, or that it was so
severely blended with a stronger transition that the limit is not
useful.}
\label{tab:ewtab}
\end{deluxetable}

\newpage
\begin{deluxetable}{lccc}
\tablenum{2}

\tabletypesize{\footnotesize}
\tablewidth{0pt}
\tablecaption{Voigt Profile Fits for Selected Transitions}
\tablehead{
\colhead{Transition} &
\colhead{$\log N$} &
\colhead{$b$ [{\kms}]} &
\colhead{$v$ [{\kms}]}
}
\startdata
\multicolumn{4}{c}{\sc $z=0.8181$ System}\\
\hline
{\MgII}  & $12.04\pm0.03$  & $2.1\pm0.4$ & $0$ \\
{\CII}   & $13.05\pm0.05$ & $6.6\pm1.8$ & $0$ \\
{\SiIII} & $12.42\pm0.10$ & $6.5\pm3.0$ & $0$ \\
{\SiIV}  & $12.47\pm0.07$ & $12.9\pm2.7$ & $0$\\
{\CIV}   & $13.53\pm0.02$ & $10.4\pm0.5$ & $0$\\
\hline
\multicolumn{4}{c}{\sc $z=0.9056$ System}\\
\hline
{\MgII}  & $12.47\pm0.01$  & $2.77\pm0.10$ & $0$ \\
{\CII}   & $13.59\pm0.04$  & $7.7\pm1.1$ & $0$ \\
{\SiIII} & $12.44\pm0.12$  & $4.3\pm1.6$ & $0$ \\
{\SiIV}  & $12.84\pm0.11$  & $9.7\pm2.8$ & $0$ \\
{\CIV}   & $13.95\pm0.23$  & $5.5\pm1.8$ & $0$ \\
         & $13.94\pm0.09$  & $13.9\pm1.3$ & $15$ \\
\hline
\multicolumn{4}{c}{\sc $z=0.6534$ System}\\
\hline
{\MgII}  & $11.8\pm0.1$      & $4.0\pm2.0$    &  $0$ \\
{\CIV}   & $\sim13.7$      & $\sim13$   &  $0$ \\
         & $\sim13.9$      & $\sim9$    &  $24$ \\
         & $\sim13.8$      & $\sim14$   &  $54$ \\
\hline
\enddata
\vglue -0.05in

\tablecomments{Velocities are offsets from the {\MgII} cloud.
For the resonant doublet transitions, the Voigt profile fits were generally constrained by
both members, however, {\SiIV} for the $z=0.9056$ system was measured from just the 1403~{\AA}
transition.  Unique Voigt profile fits could not be obtained for the $z=0.6534$ system
because of blending.  The three cloud fit given for {\CIV} provides an adequate fit
to the doublet, assuming that one component is centered on the {\MgII} cloud.}

\label{tab:vptab}
\end{deluxetable}

\newpage
\begin{deluxetable}{rcccrcc}
\tablenum{3}

\tabletypesize{\footnotesize}
\tablewidth{0pt}
\tablecaption{Guide to Model Phases for the Three Systems}
\tablehead{
\colhead{$z_{abs}$} & 
\colhead{$v$~{\kms}} &
\colhead{$\log U$} &
\colhead{$Z/Z_{\odot}$} &
\colhead{Size ({\rm kpc})} &
\colhead{$b$~({\kms})} &
\colhead{$N({\HI})$} \\
}
\startdata
\hline
0.8182 & $0$ & $-6.0$ to $-4.0$ & $\ge2$ & $10^{-6}$ to $10^{-4}$ & $2$ & $15.6$ to $15.9$\\ 
             & $0$ & $-2.2$ to $-1.8$ & $\ge1$ & $0.05$ to $0.15$ & $10$ & $14.4$ to $15.4$\\
\hline 
\hline 
0.9056 & $0$ & $-3.0$ to $-2.7$ & $\ge1$ & $0.03$ to $0.1$ & $3$ & $15.7$ \\
             & $0$ & $-1.8$ to $-1.5$ & $\ge1$ & $0.4$ to $1$ & $6$ & $14.1$ to $14.4$\\
             & $15$ & $-1.9$ to $-1.8$ & $\ge1$ & $0.4$ to $0.5$ & $14$ & $14.4$ to $14.5$\\ 
\hline
\hline 
0.6534 & $0$  & $-4.0$ to $-3.0$ & $0.03$ to $1$ & $0.0002$ to $1$ & $4$ & $15.1$ to $16.7$\\
             & $0$  & $-2.5$ to $-2.4$ & $0.1$ to $1$ & $2$ to $4$ & $13$ & $14.9$ to $16.2$\\ 
             & $24$ & $-2.2$ to $-2.1$ & $0.1$ to $1$ & $2$ to $4$ & $9$ & $14.4$ to $16.2$\\
             & $54$ & $-2.2$ to $-2.0$ & $0.1$ to $1$ & $2$ to $4$ & $14$ & $15.0$ to $16.5$\\

\hline
\hline
\enddata
\vglue -0.05in

\tablecomments{This table is intended to be a concise summary of results for the four systems.
The models that it presents are consistent with the HIRES/Keck and STIS/HST data;
they are indicative but not unique.  Abundance patterns were assume to be solar, but
this is not a unique solution, i.e.
a different abundance pattern is possible for different choices of
the other parameters.  Doppler parameters are listed for {\MgII} in the case of the low ionization
{\MgII} cloud components (first entries for each system), and for {\CIV} for all other
components.}
\label{tab:tabmod}
\end{deluxetable}

\newpage
\begin{deluxetable}{lrcclrccccccccc}
\tablenum{4}

\tabletypesize{\scriptsize}
\rotate
\tablewidth{0pt}
\tablecaption{Cloud Properties for Sample Models}
\tablehead{
\colhead{} &
\colhead{$v$}  &
\colhead{$Z$} &
\colhead{$\log U$} &
\colhead{$n_H$} &
\colhead{size} &
\colhead{$T$} &
\colhead{$N_{\rm tot}({\rm H})$} &
\colhead{$N({\HI})$} &
\colhead{$N({\MgII})$} &
\colhead{$N({\SiIV})$} &
\colhead{$N({\CIV})$} &
\colhead{$b({\rm H})$} &
\colhead{$b({\rm Mg})$} &
\colhead{$b({\rm C})$} \\
\colhead{} &
\colhead{[{\kms}]} &
\colhead{[$Z_{\odot}$]} &
\colhead{} &
\colhead{[{\cc}]} &
\colhead{[pc]} &
\colhead{[K]} &
\colhead{[{\cmsq}]} &
\colhead{[{\cmsq}]} &
\colhead{[{\cmsq}]} &
\colhead{[{\cmsq}]} &
\colhead{[{\cmsq}]} &
\colhead{[{\kms}]} &
\colhead{[{\kms}]} &
\colhead{[{\kms}]}
}

\startdata
\multicolumn{15}{c}{\sc $z=0.8181$ System} \\
\hline
{\MgII}$_{\rm 1}$ & $0$  & $2.0$ & $-4.0$ & $0.06$ & $0.1$ & $4600$ &  $16.3$ & $15.3$ & $12.0$ & $9.7$  & $9.4$  &  $9$ & $2$  & $3$ \\
{\CIV}$_{\rm 1}$ & $0$   & $2.0$ & $-2.0$ & $0.0006$ & $80$ & $8000$ & $17.2$ & $14.1$ & $10.7$ & $12.3$ & $13.5$ & $15$ & $10$ & $10$ \\

\hline
\multicolumn{15}{c}{\sc $z=0.9055$ System} \\
\hline
{\MgII}$_{\rm 1}$ & $0$  & $0.0$ & $-2.7$ & $0.003$ & $150$ & $9000$ &  $18.1$ & $15.8$ & $12.5$ & $13.1$  & $13.6$  &  $12$ & $3$  & $4$ \\
{\CIV}$_{\rm 1}$ & $0$   & $0.0$ & $-1.5$ & $0.0002$ & $1500$ & $14000$ & $18.0$ & $14.2$ & $9.1$ & $11.4$ & $14.0$ & $17$ & $4$ & $6$ \\
{\CIV}$_{\rm 2}$ & $15$   & $0.0$ & $-1.8$ & $0.0004$ & $600$ & $14000$ & $17.9$ & $14.4$ & $10.2$ & $12.2$ & $14.0$ & $20$ & $14$ & $14$ \\

\hline
\multicolumn{15}{c}{\sc $z=0.6534$ System} \\
\hline
{\MgII}$_{\rm 1}$ & $0$  & $0.1$ & $-4.0$ & $0.06$ & $2$    & $11000$ &  $17.5$ & $16.3$ & $11.8$ & $9.9$  & $9.6$  &  $14$ & $4$  & $5$ \\
{\CIV}$_{\rm 1}$ & $0$   & $0.1$ & $-2.5$ & $0.002$ & $1500$ & $19000$ & $19.0$ & $16.1$ & $11.8$ & $12.8$ & $13.7$ & $21$ & $12$ & $13$ \\
{\CIV}$_{\rm 2}$ & $24$   & $0.1$ & $-2.2$ & $0.001$ & $3200$ & $22000$ & $19.0$ & $15.8$ & $11.2$ & $12.6$ & $13.9$ & $20$ & $8$ & $9$ \\
{\CIV}$_{\rm 3}$ & $54$   & $0.1$ & $-2.2$ & $0.001$ & $2500$ & $22000$ & $18.9$ & $15.7$ & $11.1$ & $12.5$ & $13.8$ & $23$ & $14$ & $14$ \\

\hline
\enddata
\vglue -0.05in

\tablecomments{
\baselineskip=0.7\baselineskip
Column densities are listed in logarithmic units.}

\label{tab:tab4}
\end{deluxetable}


\begin{thebibliography}{XXX}

\bibitem[Bergeron and Boiss\'{e} (1991)]{bb91}
Bergeron, J., \& Boiss\'{e}, P. 1991, A\&A, 243, 344

\bibitem[Bruzual and Charlot (1993)]{bc93}
Bruzual, A. G., \& Charlot, S.
 1993, \apj, 405, 538

\bibitem[Charlton \etal (2000)]{low1634}
Charlton, J. C., Mellon, R. R., Rigby, J. R., \& Churchill, C. W. 2000,
ApJ, 545, 635

\bibitem[Churchill (1997)]{thesis}
Churchill, C. W. 1997, Ph.D. Thesis, University of California, Santa
Cruz

\bibitem[Churchill and Charlton (1999)]{q1206}
Churchill, C. W., \& Charlton, J. C. 1999, AJ, 118, 59

\bibitem[Churchill \etal (2000a)]{archive1}
Churchill, C. W., Mellon, R. R., Charlton, J. C., Jannuzi, B. T.,
Kirhakos, S., Steidel, C. C., \& Schneider, D. P. 2000a, ApJS,
130, 91

\bibitem[Churchill \etal (2000b)]{archive2}
Churchill, C. W., Mellon, R. R., Charlton, J. C., Jannuzi, B. T.,
Kirhakos, S., Steidel, C. C., \& Schneider, D. P. 2000, ApJ,
543, 577

\bibitem[Churchill \etal (1999)]{weak1}
Churchill, C. W., Rigby, J. R., Charlton, J. C., \& Vogt, S. S. 1999,
ApJS, 120, 51

\bibitem[Churchill and Vogt (2001)]{strong1}
Churchill, C. W. \& Vogt, S. S. 2001, AJ, 122, 679

\bibitem[Churchill, Vogt, and Charlton (2003)]{cvc02}
Churchill, C. W., Vogt, S.  S., \& Charlton, J. C.  2003, ApJ, in
press

\bibitem[Ding \etal (2002)]{ding1634}
Ding, J., Charlton, J. C., Zonak, S. G., \& Churchill, C. W. 2002,
ApJ, submitted

\bibitem[Ding \etal (2003)]{ding1206}
Ding, J., Charlton, J. C., Churchill, C. W., \& Palma, C. 2003,
ApJ, in preparation

\bibitem[Ferland (1996)]{cloudy}
Ferland, G. J. 1996, Hazy, University of Kentucky Internal Report

\bibitem[Haardt and Madau (1996)]{haardtmadau96}
         Haardt, F., and Madau, P. 1996, \apj, 461, 20

\bibitem[Hurwitz, Jelinsky, and Dixon (1997)]{starburst}
	Hurwitz, M., Jelinsky, P., \& Dixon, W. V. D. 1997, \apj, 481, L31

\bibitem[Lauroesch \etal (1996)]{jtl96}
         Lauroesch, J. T., Truran, J. W., Welty, D. E., and 
         York, D. G. 1996, \pasp, 108, 641

\bibitem[Rigby \etal (2002)]{weak2}
Rigby, J. R., Charlton, J. C., \& Churchill, C. W. 2002, ApJ, 565, 743

\bibitem[Steidel (1995)]{s95}
Steidel, C. C. 1995, in QSO Absorption Lines, ed. G. Meylan (Garching:
Springer--Verlag), 139

\bibitem[Steidel, Dickinson, and Persson (1994)]{sdp94}
Steidel, C. C., Dickinson, M. \& Persson, E. 1994, ApJ, 437, L75 

\bibitem[Sutherland and Dopita (1993)]{sutherland93}
Sutherland, R.S., \& Dopita, M.A., 1993, ApJS, 88, 253

\bibitem[Vogt \etal (1994)]{vogt94}
Vogt, S. S., \etal 1994, in Proceedings of the SPIE, 2128, 326

\bibitem[Zonak \etal (2002)]{zonak1634}
Zonak, S. G., Charlton, J. C., Ding, J., Churchill, C. W., \& Palma, C.
2002, ApJ, submitted

\end{thebibliography}
\end{document}